\documentclass[11pt]{article}
\usepackage{amsmath,amssymb}
\usepackage[dvips]{graphicx}

\begin{document}

\begin{titlepage}

\begin{center}
\hfill OU-HET 476 \\
\hfill UOSTP-04103 \\
\hfill SNUTP-04012\\
\hfill hep-th/0407253 \vspace{2cm}

{\Large\bf A Geometric Look on the Microstates 
of Supertubes} \vspace{1.5cm}

{\large Dongsu Bak$^a$, Yoshifumi Hyakutake$^b$, Seok Kim$^c$ and
Nobuyoshi Ohta$^b$}

\vspace{.7cm}

$^a${\it Physics Department, University of Seoul, Seoul 130-743,
Korea\\
{\tt \small dsbak@mach.uos.ac.kr}
\\[.4cm]}
$^b${\it Department of Physics, Osaka University,
Toyonaka, Osaka 560-0043, Japan\\
{\tt\small hyaku@het.phys.sci.osaka-u.ac.jp,
ohta@phys.sci.osaka-u.ac.jp}
\\[.4cm]}
$^c${\it School of Physics, Seoul National University,
Seoul 151-747, Korea\\
{\tt \small calaf2@snu.ac.kr}
\\[.4cm]} 

\end{center}
\vspace{1.5cm}

\begin{quote}

We give a geometric interpretation of the entropy of the
supertubes with fixed conserved charges and angular momenta in
two different approaches using the DBI action and the supermembrane
theory. By counting the geometrically allowed microstates, it is
shown that both the methods give consistent result on the
entropy. In doing so, we make the connection to the gravity microstates
clear.

\end{quote}

\end{titlepage}
\setcounter{page}{2}

\section{Introduction}

Supertubes are tubular shaped bound states of D0-branes,
fundamental strings (F1) and D2-branes~\cite{mateos}-\cite{WH}.
While having translational invariance in the axial direction along
which the F1 strings are stretched, the cross sectional shape of
the supertubes may be arbitrary in the eight transverse
dimensions. As shown in Ref.~\cite{karch}, the cross sectional
shape could be either open and stretched to infinity or closed but
here we would like to focus on the closed cases.

Let us begin our discussion with the cases where the cross
sectional curve lies in $x^1$ and $x^2$ plane. The supertube then
carries an angular momentum density $J=J_{12}$ proportional to the cross
sectional area. For the fixed conserved charges, the moduli space
of supertubes is consisting of the geometric fluctuations of the
cross sectional shape~\cite{BHO}. Since the angular momentum is
fixed, the fluctuation of the curve has to be area preserving. The
length $L$ of the cross sectional curve is further limited by
$\sqrt{Q_0 Q_1}$ where $Q_0$ and $Q_1$
denote lineal D0 density in the axis direction and F1 charges divided by
$2\pi$ respectively.
Thus one has the restriction of the length by
$\sqrt{J/T_2} \le\ L \ \le \sqrt{Q_0 Q_1/T_2}$,
where $T_2$ is the D2-brane tension. This space of arbitrary
fluctuation of the curve forms an infinite
dimensional moduli space. For given curve, the magnetic field
representing the density of D0 may be arbitrary with total number
of D0-branes fixed. Moreover the shape may fluctuate into the six
more transverse directions. Hence eight arbitrary bosonic
functional fluctuations are involved as the moduli deformation.
Since the supertubes involve a nonvanishing electric field and
linear momentum densities fixed by the shape of the curve, the
above moduli space is not a configuration space but a phase space.

The supertubes allow corresponding supergravity
solutions~\cite{emparan,ng} of an arbitrary cross sectional shape
and arbitrary density of D0-brane as a function of the world-volume
coordinate $\phi$ of the curve direction. Therefore the solution
involves the same number of arbitrary functions of bosonic
degrees. The geometry is nonsingular everywhere as argued in
Ref.~\cite{luninmal} in the U-dual picture. The solution does not
have a horizon either. The recently emerging picture is that such
a regular, no-horizon solution corresponds to distinguishable
gravity microstates represented by supergravity
fields~\cite{mathur}. When all the conserved charges and certain
asymptotic conditions on the geometries are fixed, the logarithm
of the number of above microstates is the entropy of the gravity
system with certain macroscopic parameters fixed. In case of
supergravity supertubes, we are interested in all the
supersymmetric solutions with fixed energy, D0 and F1 charges and
the angular momenta. The system may have many components of
angular momenta of $SO(8)$ since the system involves eight
transverse dimensions. We fix here the four independent Cartan
elements of $SO(8)$.

This solution space with all the macroscopic conserved quantities
fixed, forms a moduli space of the supergravity supertubes. As we
see in the case of DBI description of supertubes, this space must
be a phase space instead of a configuration space. Since the phase
moduli space involves arbitrary functions, it is an infinite
dimensional space. Hence its volume divided by $(2\pi \hbar)^{\rm
dim}$ is either zero or infinity. Consequently it requires at
least a regularization procedure. By quantization, the above
problem may be avoided but this will not be that simple since a
direct quantization of gravity is not well defined as we know very
well.

But the two sides of the system has a striking similarity in its
geometric nature within the moduli space. Namely the bosonic
sector in each side may be visualized as a geometric shape. The
cross sectional shape of the DBI description has a gravity
counterpart of the supertube shape. The moduli space of a
supertube consists of the shape fluctuations and again each has
its own counterpart in the supergravity side.

In fact, there is a regime where both descriptions may have their
validity. Note that the radius squared of the circular supertubes
is given by
\begin{equation}
R^2 = 2\pi g_s \ell_s^2\, {N_0 N_1}\, {\ell_s\over L_z}\,,
\label{areazero}
\end{equation}
where $N_0=L_z Q_0$ and $N_1= 2\pi Q_1$
are the numbers of D0 and F1 and $L_z$ is
the size of the compactified circle along which the axis direction
of the supertube is wrapped. This is a new length scale introduced
by supertubes and this estimation of the supertube size is valid
unless $JL_z/(Q_0 Q_1) \ll 1$.
The cross sectional area is quantized, which is
related to the quantization of the angular momentum. Considering
the case where $\gamma=L_z/(2\pi \ell_s)$ is of order one, the
validity of supergravity description requires that
\begin{equation}
g_s \, {N_0 N_1} \, \gg \, 1\,,
\end{equation}
by $R \, \gg \, \ell_s$.

Since the energy of the supertubes are given by
\begin{equation}
M= \left({1\over 2\pi \ell_s^2} N_1 + {N_0\over g_s \ell_s L_z }
\right) L_z ={1\over g_s \ell_s} \left( g_s N_1 \gamma + {N_0
}\right)\,,
\end{equation}
the Schwarzschild radius $R_S= (Mg_s^2 \ell_s^8/L_z )^{1\over 6}$ is
\begin{equation}
R_S=\ell_s \left[ \left(g_s\, N_1 + N_0/\gamma \right)g_s/2\pi
\right]^{1\over 6}\,.
\end{equation}
Thus, for $g_s \, {N_0 N_1} \, \gg \, 1$, $R \, \gg\, R_S$, which
may explain the regularity of the supertube solutions.

On the other hand, the DBI description has its validity in the
decoupling limit of $\ell_s\,\rightarrow\, 0 $ and
$g_s\,\rightarrow\, 0$. Thus the overlapping region of the
validity is given by the open-string decoupling limit,
\begin{equation}
\ell_s\,\rightarrow\, 0,\ \ \ g_s\,\rightarrow\, 0\,,
\end{equation}
while keeping the combination $g_s N_0 N_1$ large.
Thus we have here the gravity and the supertube field theory
correspondence in the overlapping regime of the validity. The
decoupled field theory is the world-volume field theory of
supertubes. In this limit, the field
theory obtained by expanding DBI theory around the circular
supertube background is eventually described by a peculiar 2+1
dimensional (noncommutative) Yang-Mills theory in the decoupling
limit where $\ell_s\, \rightarrow\, 0 $, which is equivalent to
the matrix theory in a circular supertube
background~\cite{bl,karch,jabbari}.

In this note, we would like to count the entropy of the geometries
using the gravity/field theory correspondence and the structure of
the phase moduli obtained in Ref.~\cite{BHO}. We shall be using
basically the DBI action to count the degeneracy the states. This
problem is in some sense already treated in Ref.~\cite{marolf},
but the perspective and the emphasis on the geometric nature are
the main differences.

We would like to first make it clear that we are basically
counting the geometric fluctuations in the sense that, even in the
field theory, we are counting the freedom of the shape fluctuation
including other accompanying bosonic and fermionic degrees. This
simpler version of the account of entropy using the shape
fluctuation is presented first. The full derivation of the entropy
in the decoupling limit is done via two different  methods. One is the
description of DBI action. Here we identify the infinite
dimensional fermionic moduli space and count the entropy including
all the fermionic fluctuations. In this case we use
the near circular condition $q=(Q_0 Q_1-J)/J \, \ll \, 1$
to simplify the calculation. The others are via the M-theory
description of the M2-brane. In this M2 brane picture, we find that
the near circular condition is not necessary for the counting.

As stated earlier, the quantization is necessary to get the
correct expression and in this sense the decoupling limit is
essential. At the end of the day, the
decoupled theory in the same limit should lead to a unique theory
in any paths.

The paper is organized as follows. In Sections 2 and 3, we
review the phase moduli space of supertubes and count the cross
sectional shape fluctuations. In Section 4, we explain the
fermionic part of the moduli space using DBI description and count
the full degeneracy in the near circular limit. In Sections 5,
we count again the entropy using the M-theory.
Section~6 is devoted to conclusions.

\section{BPS Equations and Conserved Charges for a Closed Supertube}
\label{sec2}

A tubular D2-brane with electric and magnetic fluxes on the
world-volume becomes a closed supertube if it satisfies suitable
BPS conditions. We first review these BPS equations and conserved
charges for a closed supertube. This also serves to establish our
notations.

The tubular D2-brane is embedded in the 10-dimensional flat space-time,
and the world-volume is parametrized by $(t,\phi,z)$. The pullback
metric and the field strength on the D2-brane is written as
\begin{alignat}{3}
&ds_{\text{pb}}^2 = -(1-|\dot{\vec{x}}|^2)dt^2 + |\vec{x}'|^2
d\phi^2 +2\dot{\vec{x}} \cdot \vec{x}'dtd\phi + dz^2,
\\
&F = E dt \wedge dz + B dz \wedge d\phi. \notag
\end{alignat}
Here $\vec{x} = (x^1,\cdots,x^8)$, $\dot{x} \equiv \frac{\partial
x}{\partial t}$ and $x' \equiv \frac{\partial x}{\partial \phi}$.
The cross section of the D2-brane is expressed as an arbitrary
loop in $\mathbb{R}^8$. The angle $\phi$ ($-\pi \leq \phi \leq
\pi$) represents the direction along the loop and $z$ lies in the
transverse direction. Thus we are considering only the configurations
with the translational invariance in the $z$ direction.

The bosonic part of the D2-brane DBI action is evaluated as
\begin{alignat}{3}
\label{eq:actD2}
S 
= - T_2 \!\int\! dt d\phi dz \sqrt{ (1 - |\dot{\vec{x}}|^2)
(|\vec{x}'|^2 + \lambda^2 B^2) - \lambda^2 E^2 |\vec{x}'|^2 +
(\dot{\vec{x}} \cdot \vec{x}')^2 - 2 \lambda^2 EB \dot{\vec{x}}
\cdot \vec{x}' },
\end{alignat}
where a D2-brane tension $T_2$ and $\lambda$ are written as $T_2 =
\frac{1}{(2\pi)^2\ell_s^3 g_s}$ and $\lambda = 2\pi \ell_s^2$,
respectively, in terms of the string length $\ell_s$ and coupling
$g_s$. Canonical momenta $p_i$ and $\Pi$ conjugate to $x^i$ and
$A_z$ are written as
\begin{alignat}{3}
\label{can} p_i &= - \frac{T_2^2}{\mathcal{L}}\big\{ \dot{x}_i
(|\vec{x}'|^2 + \lambda^2 B^2) - (\dot{\vec{x}} \cdot
\vec{x}'){x_i}' + \lambda^2 EB {x_i}' \big\}, \nonumber
\\
\Pi &= - \frac{T_2^2}{\mathcal{L}}\big\{ \lambda^2 E |\vec{x}'|^2
+ \lambda^2 B \dot{\vec{x}} \cdot \vec{x}' \big\},
\end{alignat}
where $\mathcal{L}$ is the Lagrangian density. The Hamiltonian
density is given by
\begin{alignat}{3}
\mathcal{H} &= \sqrt{T_2^2|\vec{x}'|^2 + T_2^2\lambda^2 B^2 +
|\vec{p}|^2 + \frac{\Pi^2}{\lambda^2}} \notag
\\
&= \sqrt{ \bigg( \frac{\Pi}{\lambda} + T_2 \lambda B \bigg)^2 +
\bigg( T_2|\vec{x}'| - \frac{\Pi B}{|\vec{x}'|} \bigg)^2 + \bigg(
|\vec{p}|^2 - \frac{\Pi^2 B^2}{|\vec{x}'|^2} \bigg) } \notag
\\
&\ge T \Pi + T_0 \frac{B}{2\pi}, \label{ineq1}
\end{alignat}
where $T=\frac{1}{2\pi\ell_s^2}$ is the tension of the fundamental
string and $T_0 = \frac{1}{\ell_s g_s}$ is the mass of a D0-brane.
It can be verified that the third term in the square root in the second
line is non-negative and vanishes when $\dot{\vec{x}} \propto \vec{x}'$,
which is equivalent to $\dot{\vec{x}} = 0$ by suitable reparametrization
of $\phi$. Thus the Hamiltonian density is bounded from below by the
mass density of fundamental strings and D0-branes, which precisely
matches with the energy of the supertube. The equality in
(\ref{ineq1}) is saturated when
\begin{alignat}{3}
\Pi B = T_2 |\vec{x}'|^2 \equiv T_2 \Big(\frac{ds}{d\phi}\Big)^2,
\qquad \dot{\vec{x}} = 0, \label{eq:BPS}
\end{alignat}
where $ds^2 = d\vec{x} \cdot d\vec{x}$ is the line element of
$\mathbb{R}^8$. These are the BPS conditions which must be
satisfied by all supertubes.

The closed supertube carries two charges, which corresponds to
those of fundamental strings and D0-branes, and angular momenta.
Defining
\begin{alignat}{3}
Q_1 = \frac{1}{2\pi} \int_{-\pi}^\pi d\phi\,\, \Pi, \qquad Q_0 =
\frac{1}{2\pi} \int_{-\pi}^\pi d\phi\,\, B, \qquad
\label{eq:elemag}
\end{alignat}
we find that $2\pi Q_1 \in \mathbb{Z}$ is the number of
fundamental strings dissolved in the D2-brane, and $Q_0 L_z \in \mathbb{Z}$
is that of D0-branes. The angular momenta are given as
\begin{alignat}{3}
L^{ij} &= \frac{1}{2\pi} \oint_{-\pi}^\pi d\phi \,(x^i p^j - x^j
p^i) = \frac{T_2}{\pi} \int dx^i \wedge dx^j, \label{eq:angmomc}
\end{alignat}
where $i,j=1,\cdots,8$. With the aid of the BPS equations
(\ref{eq:BPS}), the canonical momenta are expressed as $p^i = T_2
{x^i}'$. The last equality in (\ref{eq:angmomc}) is obtained by
using this relation. Thus, when we fix the angular momenta, the
area made by projecting the loop of the supertube onto
$(i,j)$-plane should be preserved during the deformation in the
flat directions.

The flat directions of supertubes of general shape with fixed
fundamental string charge $Q_1$, D0-brane charge $Q_0$ and angular
momentum $J$ are of our interest. They make the moduli space of
the supertubes and are related to the number of gravity microstates.
It was shown~\cite{BHO} that the perimeter $2\pi L$ of the supertubes
is restricted by
\begin{eqnarray}
\sqrt{J/T_2} \leq L \leq \sqrt{Q_0 Q_1/T_2}.
\label{range}
\end{eqnarray}
The number of microstates allowed by this bound is the problem we
are going to discuss in this paper.

\section{A First Look on the Microstates of Supertubes}

In order to get the idea how to count the microstates of the
supertubes, let us first discuss the bosonic
fluctuations $\epsilon(\phi)$, $a(\phi)$ and
$b(\phi)$ around the circular background,
\begin{alignat}{3}
\zeta &= x_1+i x_2= R(1+\epsilon) e^{i\phi}, \qquad \Pi = Q_1 (1+a), \qquad B
= Q_0(1+b) \label{eq:flu1}
\end{alignat}
where $\epsilon$, $a$ and $b$ are real. We consider the landscape
of vacua having ${Q}_0 {Q}_1 > T_2 R^2$ with the area ${\cal A}=\pi R^2$
fixed. Introducing $q$ defined by
\begin{alignat}{3}
Q_0 Q_1 &= T_2 R^2(1+q), \label{eq:flu}
\end{alignat}
we focus on the case of $q \ll 1$. The fluctuation can be
expanded as\footnote{As we shall see later on, $a$ and $b$ are mixed
with the $\epsilon$ fluctuation. But for simplicity, we ignore this
complication here. See also Ref.~\cite{BHO} for the detailed classical
description of this mixing.}
\begin{alignat}{3}
& \epsilon(\phi) = \sum_{n \in \mathbb{Z}} \epsilon_n e^{in\phi},
\qquad a(\phi) = \sum_{n \in \mathbb{Z}} a_n e^{in\phi}, \qquad
b(\phi) = \sum_{n \in \mathbb{Z}} b_n e^{in\phi}.
\end{alignat}
Here we should impose reality conditions $\epsilon_{-n} =
\epsilon_n^\dagger, a_{-n} = a_n^\dagger$ and $b_{-n} =
b_n^\dagger$.

The area given by
\begin{alignat}{3}
& {\cal A} = {1\over 4} \int_0^{2\pi} {\rm Im} (\zeta d\zeta^\dagger
-\zeta^\dagger d\zeta)
\end{alignat}
is evaluated as
\begin{alignat}{3}
& {\cal A} = \pi R^2\left(1+ 2\epsilon_0+\sum_{n \in \mathbb{Z}}
|\epsilon_n|^2\right)\,.
\end{alignat}
The conservation of the angular momentum implies that
\begin{alignat}{3}
& 2\epsilon_0=-\sum_{n \in \mathbb{Z}} |\epsilon_n|^2\,.
\label{area1}
\end{alignat}

The length $2\pi L$ of the curve
\begin{alignat}{3}
& 2\pi L = \int_0^{2\pi} |d\zeta|=R\int_0^{2\pi}d\phi
\sqrt{(1+\epsilon)^2+(\epsilon')^2}
\end{alignat}
may be expanded as
\begin{alignat}{3}
2\pi L = R\int_0^{2\pi}d\phi
\left(1+{(2\epsilon+\epsilon^2+(\epsilon')^2) \over
2}-{\epsilon^2\over 2}+\cdots\right) = 2\pi
R\left(1+\epsilon_0+\sum_{n \in \mathbb{Z}} {n^2\over
2}\,|\epsilon_n|^2+\cdots\right)\,. \label{length1}
\end{alignat}
The condition (\ref{area1}) may be used to eliminate $\epsilon_0$
from (\ref{length1}) and we get
\begin{alignat}{3}
& 2\pi L = 2\pi R\left(1+ \sum_{n > 1}
(n^2-1)\,|\epsilon_n|^2+\cdots\right)\,.
\end{alignat}
Here $\epsilon_n$ for $n > 0$ are our phase space variables and
$\epsilon_0$ is constrained by the condition (\ref{area1}). We are
not interested in the translational mode $\epsilon_1$.

Since from (\ref{range})
\begin{alignat}{3}
& R \ \le\ L \ \le \ \sqrt{Q_0 Q_1/T_2}\,,
\end{alignat}
we get, using (\ref{eq:flu}), a constraint
\begin{alignat}{3}
& \sum_{n > 1} (n^2-1)\,|\epsilon_n|^2\ \le \ q/2\,. \label{cons1}
\end{alignat}

To compute the number of states in the volume~(\ref{cons1}), let
us find out the canonical variables in the phase space. First we
define the coordinate corresponding to the radius as
$r(=R(1+\epsilon))$ which is real. From the
action~(\ref{eq:actD2}), we obtain the conjugate momentum
\begin{equation}
p_r(\phi) = T_2 r' \ , \label{mom}
\end{equation}
where use has been made of the fact that $E=1/\lambda$ for the BPS
states of our concern here and ${\cal L}= -T_2 \lambda\, B$. The
relation~(\ref{mom}) is a second class constraint and we should
make Dirac quantization. After this procedure, one finds
\begin{alignat}{3}
[r(\phi), T_2 r'(\phi')]=\frac{i}{2}
\delta(\phi-\phi')\delta(z-z')\,.
\end{alignat}
For the zero mode in the $z$ direction, the above implies that
\begin{alignat}{3}
& [\epsilon^\dagger_m, \epsilon_n] = {1 \over 4\pi T_2 R^2 L_z
n}\,\delta_{mn}\,,
\end{alignat}
where $L_z$ is the length of the supertube in the $z$ direction.
Thus $c_n$ defined by
\begin{alignat}{3}
& {\alpha\over \sqrt{n}}\,\, c^\dagger_n =\epsilon_n \,,
\end{alignat}
with $\alpha^2=1/(4\pi T_2 R^2 L_z)$ satisfies the commutation
relation
\begin{alignat}{3}
& [c_m, c^\dagger_n] = \delta_{mn}\,.
\end{alignat}
In terms of these variables, the constraint (\ref{cons1}) becomes
\begin{alignat}{3}
& \sum^\infty_{n = 2} \Big(n-\frac{1}{n}\Big)\,|c_n|^2 \ \le \
{q\over 2\alpha^2}\,.
\end{alignat}
Quantum mechanically, this condition is interpreted as
\begin{alignat}{3}
& \sum^{\infty}_{n = 2} \Big(n-\frac{1}{n}\Big) N_n\, \ \le \
{q\over 2 \alpha^2}\equiv s\,, \label{conin}
\end{alignat}
where the number operator $N_n$ is defined by $c^\dagger_n c_n$.

Our task is now to evaluate the number of states restricted by
(\ref{conin}). For large $n$, the $\frac1n$ in the bracket may be
ignored, and the problem reduces to the well-known case of
counting string states. It is given by
\begin{alignat}{3}
& {\cal V}={\sqrt{2}\over 4\pi \sqrt{s}} e^{\pi\,\sqrt{{2\over
3}s}}\,. \label{hagedorn}
\end{alignat}

To get this, let us consider the following quantity~\cite{gsw}
\begin{alignat}{3}
G(w) = \sum_{n=0}^\infty d_n w^{nN_n} = {1\over \prod^\infty_{n=1}
(1-w^n) } \equiv [f(w)]^{-1}\,.
\end{alignat}
This is related to the Dedekind eta function
\begin{equation}
\eta(\tau) = e^{i\pi\tau/12}\prod_{n=1}^\infty (1-e^{2\pi i n\tau}),
\label{eta}
\end{equation}
which has the modular transformation formula
\begin{equation}
\eta(-1/\tau)=(-i\tau)^{1/2} \eta(\tau).
\label{modular}
\end{equation}
Applied to $f(w)$, this gives the Hardy-Ramanujan formula
\begin{equation}
f(w)=\left(\frac{-2\pi}{\log w}\right)^{1/2} w^{-1/2} \tilde w^{1/12}
f(\tilde w^2),
\end{equation}
where
\begin{equation}
\tilde w = e^{2\pi^2/\log w}.
\end{equation}
One can then deduce the asymptotic formula for $w \to 1$ (or $\tilde w\to 0$)
\begin{equation}
f(w) \sim \left(\frac{-2\pi}{\log w}\right)^{1/2}
\exp\left(\frac{\pi^2}{6\log w}\right). \label{asym}
\end{equation}
The degeneracy is obtained by
\begin{equation}
d_n = \oint \frac{G(w)}{w^{n+1}} \frac{dw}{2\pi i}. \label{deg}
\end{equation}
Using the asymptotic expansion of $f(w)$, this can be estimated
for large $n$ by a saddle point evaluation. $G(w)$ grows rapidly
for $w\to 1$, while if $n$ is very large, $w^{n+1}$ is very small
for $w<1$. There is a sharp saddle point for $w$ near 1. The integrand
(for the integration variable $\log w$)
\begin{equation}
\exp\left( -\frac{\pi^2}{6\log w} - n \log w\right),
\end{equation}
is stationary for $\log w \sim -\pi/\sqrt{6 n}$. Evaluating
(\ref{deg}) around this saddle point, we get~\cite{hage}
\begin{alignat}{3}
d_n = {1\over 4\sqrt{3}\,n} e^{\pi\,\sqrt{{2\over 3}n} }\,,
\end{alignat}
in the large $n$. Integration of $d_n$ up to $s$ gives ${\cal V}$ in
(\ref{hagedorn}).

Thus the entropy becomes
\begin{alignat}{3}
& S=\ln\, {\cal V} =\pi \sqrt{q\over 3 \alpha^2} = \pi
\sqrt{{4\pi\over 3} \, L_z\, (Q_0 Q_1-J)}\,. \label{bos}
\end{alignat}
We note that this is the entropy from a single fluctuating boson
around the supertube. In what follows we are going to extend this
to supertubes with other modes.

\section{Supertube solutions in DBI action and entropy}

In this section we find exact BPS supertube solutions including fermion
backgrounds using the DBI action of D2. The BPS solutions
preserve $\frac{1}{4}$ supersymmetry. One may identify
`fermionic flat directions' of the classical solutions. We then give
the quantization rules for the flat modes and we count the contributions
to the entropy from fermions as well as bosons.

\subsection{The solutions}

We start by summarizing the supersymmetric DBI theory for the
D2  in the notations and conventions of~\cite{APS}. This
action has  gauge invariances coming from worldvolume
diffeomorphism and the local kappa symmetry. Since the full
gauge invariant action is complicated, we start from the action
with gauge fixed kappa symmetry:
\begin{equation}
\label{inv-action} S=-\int
d^3\sigma\sqrt{-\det[g_{\mu\nu}+F_{\mu\nu}-
2\bar\lambda\gamma_{\mu}\partial_{\nu}\lambda+
(\bar\lambda{\bf\Gamma}^M\partial_{\mu}\lambda)
(\bar\lambda{\bf\Gamma}_M\partial_{\nu}\lambda)]},
\end{equation}
where $\mu,\nu=t,\phi,z$ are the worldvolume indices, $M$ is the
$R^{9+1}$ vector index, $g_{\mu\nu}$ is the pullback of
10-dimensional flat metric onto the world-volume,
$\gamma_\mu={\bf\Gamma}_M\frac{\partial X^M}{\partial\sigma^\mu}$
is the induced gamma matrix, and $\lambda$ is the Majorana-Weyl
fermion in the target space, where the Weyl condition is imposed
by the gauge choice for the local kappa symmetry. We use the
convention $\bar\lambda=\lambda^\dag(-i{\bf\Gamma}^0)$. As in
Section~\ref{sec2}, one should also include an overall coefficient
$T_2$, the D2-brane tension, and also replace $F$ by $2\pi\ell_s^2
F$.

For later use, we summarize our gamma matrix conventions. The
$SO(9,1)$ $32\times 32$ $\Gamma^M$ is expressed in terms of
$SO(1,1)$ $2\times 2$ gamma matrices and $SO(8)$ $16\times 16$
gamma matrices as follows:
\begin{equation}
\label{gamma} 
{\bf\Gamma}^0=(i\sigma^2)\otimes{\bf
1}_{16},\ \
{\bf\Gamma}^{z}=
\sigma^1\otimes{\bf 1}_{16},\ \ {\rm others}\ \sigma^3\otimes\Gamma^i
\ (i=1,2,\cdots, 8)\ .
\end{equation}
$\Gamma^i$'s are the $SO(8)$ spinors in a suitable representation:
We use the convention that the former/latter 8 indices act on
left/right chiral components, respectively. The last eight gamma
matrices in (\ref{gamma}) will be written as
$\vec{\bf\Gamma}=\sigma^3\otimes\vec\Gamma$, where the vector
lives in $R^8$. The chirality operator is defined as
\begin{equation}\label{chiral}
{\bf\Gamma}_{11}={\bf\Gamma}_0{\bf\Gamma}_1\cdots{\bf\Gamma}_8 {\bf\Gamma}_z =
\sigma^3\otimes\Gamma^9,
\end{equation}
with the choice $\Gamma^9=\mbox{diag}({\bf 1}_8,-{\bf 1}_8)$. The
Weyl condition on $\lambda$ is chosen to be
${\bf\Gamma}_{11}\lambda=+\lambda$.
With our convention (\ref{gamma}) and (\ref{chiral}), this chiral
$\lambda$ is written as
\begin{equation}
\label{weyl-ferm}
\lambda(\sigma)=\left[\begin{array}{c}1\\0\end{array}\right]\otimes
\left[\begin{array}{c}\psi(\sigma)\\0\end{array}\right]+
\left[\begin{array}{c}0\\1\end{array}\right]\otimes
\left[\begin{array}{c}0\\ \chi(\sigma)\end{array}\right],
\end{equation}
where $\psi,\chi$ are 8-component $SO(8)$ spinors with $\Gamma^9$
eigenvalues $\pm 1$, respectively.

Let us make the \textit{partial gauge fixing} for the world-volume
diffeomorphism
\begin{equation}
T=t,\ Z=z,\ 0\leq\phi<2\pi,\ \vec{x}(t,\phi,z)\in R^8,
\end{equation}
\textit{i.e.}, we leave unspecified one scalar field in $R^8$
tangent to $\phi$ direction. This is harmless as far as we do the
classical analysis.\footnote{However, when we consider
quantization of near-circular supertubes, we should fix this extra
gauge. There we will set this scalar equal to $\phi$.} In this
gauge, the action (\ref{inv-action}) is written as
\begin{equation}
S=-\int d^3 \sigma \sqrt{-\det {\mathcal{M}}_{\mu\nu}}\ ,
\end{equation}
where
\begin{equation}
{\mathcal{M}}_{\mu\nu}=\eta_{\mu\nu}+F_{\mu\nu}+\partial_{\mu}\vec{x}\cdot
\partial_{\nu}\vec{x}-2\bar\lambda(\tilde\gamma_{\mu}+\vec\Gamma\cdot
\partial_{\mu}\vec{x})\partial_{\nu}\lambda+
(\bar\lambda{\bf\Gamma}^M\partial_{\mu}\lambda)
(\bar\lambda{\bf\Gamma}_M\partial_{\nu}\lambda)\ .
\end{equation}
This is the final supersymmetric DBI action that we need. Note
that, again due to the partial gauge fixing,
$-\eta_{tt}=\eta_{zz}=1$ but $\eta_{\phi\phi}=0$; furthermore,
$\tilde\gamma_t={\bf\Gamma}_0$ and $\tilde\gamma_z={\bf\Gamma}_z$
but $\tilde\gamma_\phi=0$.

The solution we are looking for is independent of $t$ and $z$, so
we take $\vec{x}(\phi)$, $F=E(\phi)dt\wedge dz+B(\phi)dz\wedge
d\phi$ and $\lambda(\phi)$ and insert them into the equations of
motion. In this process, we may set all $t,z$ derivatives of
$\vec{x}$ in the Lagrangian to zero, since terms containing these
derivatives would not survive the equations of motion for
$\vec{x}$. We can then rewrite $\sqrt{-\det{\mathcal M}_{\mu\nu}}$ as
\begin{equation}
\label{determinant} \sqrt{|\vec{x}'|^2(1-E^2)+B^2+(1-E^2)
\{(\bar\lambda{\bf\Gamma}^M\lambda')^2\!-2\bar\lambda(\vec{\Gamma}\cdot
\vec{x}')\lambda'\}
+2B\{E\bar\lambda{\bf\Gamma}_0\lambda'-\bar\lambda{\bf\Gamma}_2\lambda'\}}\ ,
\end{equation}
where the prime denotes $\phi$ derivative. Variation of this
quantity in $\lambda,\vec{x}$ and $A_\mu$ yields equations of
motion. As we know that the supertube solution is obtained for
$E=1$, let us set $E=1$ after variation in these fields, which
simplifies the resulting equations drastically.

The variation of the Lagrangian in $\delta\bar\lambda$ (with
$\delta\lambda$ since it is Majorana) gives (after setting $E=1$)
\begin{equation}
\label{fermion-eom}
\delta{\mathcal L}
=-\frac{B}{\sqrt{B^2+2B\bar\lambda({\bf\Gamma}_0-{\bf\Gamma}_z)\lambda'}}
\left[\delta\bar\lambda({\bf\Gamma}_0-{\bf\Gamma}_z)\lambda'+
\bar\lambda({\bf\Gamma}_0-{\bf\Gamma}_z)\delta\lambda'\right]\,,
\end{equation}
and the equation of motion
\begin{equation}
\label{ferm-BPS} ({\bf\Gamma}_0-{\bf\Gamma}_z)\lambda'=0\,.
\end{equation}
One can easily check that the full
equations of motion are solved by $E=1$ and (\ref{ferm-BPS}) for
any functions $B(\phi)$ and $\vec{x}(\phi)$. This is a simple
generalization of the original supertube solution.

With the representation (\ref{gamma}), (\ref{ferm-BPS}) becomes
$[\sigma^3\otimes{\bf 1}_{16}]\lambda'=\lambda'$. This means that
the second term in (\ref{weyl-ferm}) should be $\phi$-independent:
\begin{equation}
\label{ferm-sol}
\lambda(\phi)=\left[\begin{array}{c}1\\0\end{array}\right]\otimes
\left[\begin{array}{c}\psi(\phi)\\0\end{array}\right]+
\left[\begin{array}{c}0\\1\end{array}\right]\otimes
\left[\begin{array}{c}0\\ \chi\end{array}\right] \ ,\ \ (\chi:{\rm
any\ constant\ spinor}).
\end{equation}
Note that the fermionic part contains arbitrary functions of
$\phi$, an 8-component $SO(8)$ spinor $\psi(\phi)$ with positive
chirality. The bosonic part has one from $B(\phi)$
plus seven gauge-invariant components from $\vec{x}(\phi)$.
Below we shall show  that $\chi\!=\!0$ for the $\frac{1}{4}$
supersymmetry. Hence the fermionic part of supertubes also involves
eight arbitrary functions of moduli fluctuation, which is expected from
the number of remaining supersymmetries.

\subsection{Supersymmetry}

We now check whether the above solution preserves $\frac{1}{4}$
supersymmetry. The 32 supersymmetry parameters $\epsilon$ of
type-IIA string theory are split into $\epsilon_\pm$ satisfying
${\bf\Gamma}_{11}\epsilon_{\pm}=\pm\epsilon_{\pm}$, respectively.
The supersymmetry transformations, combined with compensating
kappa transformation and world-volume diffeomorphism to restore
the gauge, are
\begin{eqnarray}
\label{susy} \delta\bar\lambda&=&\bar\epsilon_+ +\bar\epsilon_-
\gamma^{(z)}+ \xi^\mu\partial_\mu\bar\lambda, \nonumber\\
\delta\vec{x}&=&(\bar\epsilon_+\!-\bar\epsilon_-\gamma^{(z)})
\vec{\bf\Gamma}\lambda+\xi^\mu\partial_\mu\vec{x} , \\
\delta A_{\mu}&=&(\bar\epsilon_-\gamma^{(z)}\!-\bar\epsilon_+)
(\tilde\gamma_\mu+\vec{\bf\Gamma}\cdot\partial_{\mu}\vec{x})\lambda \nonumber\\
&&+(\frac{1}{3}\bar\epsilon_+ -\bar\epsilon_-\gamma^{(z)})
{\bf\Gamma}_M\lambda\ \bar\lambda{\bf\Gamma}^M\partial_\mu\lambda
+\xi^{\nu}\partial_\nu A_\mu+\partial_{\mu}\xi^\nu A_\nu, \nonumber
\end{eqnarray}
where
$\xi^\mu=(\bar\epsilon_-\gamma^{(z)}\!-\bar\epsilon_+)\gamma^\mu\lambda$
(with $\mu=t,z$ only in the superscript) and the $32\times 32$
matrix $\gamma^{(2)}$ in our case is (using $E\!=\!1$ and
(\ref{ferm-BPS}) to simplify the expression)
\begin{equation}\label{kappa-proj}
\gamma^{(z)}=-({\bf\Gamma}_0{\bf\Gamma}_z+{\bf\Gamma}_{11})\
\frac{\vec{x}'\cdot\vec{\bf\Gamma}-\vec{\bf\Gamma}\cdot
(\bar\lambda\vec{\bf\Gamma}\lambda')}{B}-
{\bf\Gamma}_{11}{\bf\Gamma}_0\ .
\end{equation}
Note that we have fixed the world-volume diffeomorphism only
partially, i.e., $T\!=\!t$ and $Z\!=\!z$, so in the above
definition we have two gauge-keeping parameters $\xi^t$ and
$\xi^z$ but nothing like $\xi^{\phi}$.

Let us start from $\delta\bar\lambda$. The last term is absent
after inserting our solution, so we have
\begin{equation}\label{fermion-susy}
\delta\bar\lambda=\bar\epsilon_+
+\bar\epsilon_-\left\{-(1-\sigma^3)\
\frac{\vec{x}'\cdot\vec{\bf\Gamma}-\vec{\bf\Gamma}\cdot
(\bar\lambda\vec{\bf\Gamma}\lambda')}{B}+(i\sigma^2)\right\},
\end{equation}
where we have used $\bar\epsilon_-{\bf
\Gamma}_{11}=+\bar\epsilon_-$, and for all Pauli matrices $[\
\cdot\ \otimes{\bf 1}_{16}]$ is implicit. Considering the
chiralities of $\epsilon_\pm$, we can write them in the following
form:
\begin{equation}
\epsilon_+=\left[\begin{array}{c}1\\0\end{array}\right]\otimes
\left[\begin{array}{c}\alpha_+\\0\end{array}\right]+
\left[\begin{array}{c}0\\1\end{array}\right]\otimes
\left[\begin{array}{c}0\\ \beta_+\end{array}\right],\ \
\epsilon_-=\left[\begin{array}{c}0\\1\end{array}\right]\otimes
\left[\begin{array}{c}\alpha_-\\0\end{array}\right]+
\left[\begin{array}{c}1\\0\end{array}\right]\otimes
\left[\begin{array}{c}0\\ \beta_-\end{array}\right].
\end{equation}
In order for our solution to be supersymmetric, the first term in
the curly bracket should vanish. This is true if $\beta_-=0$. The
cancellation of the remaining second term $\propto(i\sigma^2)$
with the $\bar\epsilon_-$ term requires $\alpha_+=\alpha_-$ and
$\beta_+=0$. Therefore we have $\frac{1}{4}$ of the $\epsilon$
components preserved; $\beta_+$, $\beta_-$ and
$\alpha_+\!-\!\alpha_-$ are broken.

Let us now consider the remaining transformations for bosonic
fields. Using the expression (\ref{kappa-proj}) and the conditions
$\beta_\pm\!=\!0$, $\alpha_+\!=\!\alpha_-\equiv\alpha$, the scalar
variation $\delta\vec{x}$ reduces to
\begin{equation}
\delta\vec{x}=-2i\left[\ \alpha^\dag\ \ 0\ \right]\
\vec{\Gamma}\ \left[\begin{array}{c}0\\
\chi\end{array}\right]\ .
\end{equation}
One can see from the above variation that all the supersymmetry is
broken if the constant mode $\chi$ is non-zero. Therefore, as
promised, we should set $\chi=0$ in order to have
$\frac{1}{4}$-BPS deformation.

In order to check the last transformation, we should specify the
vector potential giving rise to the field strength $F=dt\wedge dz
+B(\phi)dz\wedge d\phi$. The simplest choice may be $A^{(1)}=tdz+
B(\phi)z d\phi$. For later use, let us also consider an
alternative choice. To this end, we decompose the magnetic field
as
\begin{equation}
B_0=\frac{1}{2\pi}\oint d\phi B(\phi) \ ,\ \
b(\phi)=B(\phi)-B_0\equiv -a'(\phi),
\end{equation}
where $a(\phi)$ is a periodic function, due to the fact $\oint
d\phi\ b(\phi)=0$. Then we can choose the 1-form potential as
\begin{equation}
A^{(2)}=\{t+a(\phi)\}dz+ zB_0d\phi\ .
\end{equation}
This form would be more convenient later, when we consider
quantization. The two choices are related by a gauge
transformation. 

After inserting our supertube solution with $\chi=0$, and turning
on $\alpha$ only, the supersymmetry transformation (\ref{susy})
for gauge field becomes
\begin{eqnarray}
\label{susy-gauge}
\delta A_0^{(1)}&=&\partial_t \{2it\alpha^\dag\psi(\phi)\}, \nonumber\\
\delta A_z^{(1)}&=&0, \\
\delta A_\phi^{(1)}&=&\partial_\phi\{2it \alpha^\dag\psi(\phi)\}+
2i\alpha^\dag\psi(\phi)B(\phi)\ .\nonumber
\end{eqnarray}
If the last term in
the third line in (\ref{susy-gauge}) is absent, one can make a
compensating gauge transformation and have $\frac{1}{4}$
supersymmetry. The subtle term proportional to $\psi(\phi)B(\phi)$
can be decomposed into the `0-mode' piece plus the remainder as
\begin{equation}
\overline{\psi B}\equiv\frac{1}{2\pi}\oint d\phi\psi(\phi)B(\phi)\
,\ \ \{\psi(\phi)B(\phi)\}_r\equiv\psi(\phi)B(\phi)-\overline{\psi B}.
\end{equation}
The `remainder' piece $\{\psi(\phi)B(\phi)\}_r$ can be rewritten
as $\partial_\phi [\ \cdot\ ]$, where we have a well-defined
periodic function of $\phi$ inside the square bracket. This can
safely be compensated by a gauge transformation. The 0-mode piece
can be written as $d\{2i\phi\alpha^\dag\overline{\psi B}\}$, which
is not a gauge transformation in general. To ensure supersymmetry,
we require $\overline{\psi B}=0$, which in turn implies that
$\psi$ 0-modes are expressed in terms of $\psi$ and $B$ nonzero
modes.

\subsection{Conserved charges}

The electric displacement
$\Pi(\sigma)=\frac{\partial{\mathcal{L}}}{\partial E}$ is obtained
from (\ref{determinant}) and setting $E=1$ after differentiation:
\begin{equation}
\Pi(\sigma)=\frac{|\vec{x}'|^2}{B}-\bar\lambda{\bf\Gamma}_0\lambda'+
\frac{1}{B}\left\{(\bar\lambda{\bf\Gamma}^M\lambda')^2
-2\bar\lambda(\vec{\bf\Gamma}\cdot\vec{x}')\lambda'\right\}\ .
\end{equation}
One may show that that all the higher-order terms in the curly
bracket vanish for the supertube solution.
The electric displacement reduces to
\begin{equation}
\Pi(\sigma)
=
\frac{|\vec{x}'|^2}{B}+i\psi^\dag\psi'\,,
\end{equation}
for the supertubes.

The linear momentum density conjugate to $\vec{x}$ 
becomes
\begin{equation}
\vec{p}=\vec{x}\,'\,.
\end{equation}
Note that there is no correction from fermions. Consequently the field
angular momentum also takes the same form as in the bosonic case,
\begin{equation}
L^{ij}=\frac{1}{2\pi}\oint d\phi(x^ip^j-x^jp^i)=\frac{1}{\pi}\int
dx^i\wedge dx^j\ ,
\end{equation}
proportional to the cross section area of the tube. To obtain the
conserved total angular momentum, we should add to it the spin
angular momentum. The total angular momentum is then
\begin{equation}
J^{ij}=\frac{1}{\pi}\int dx^i\wedge dx^j - \frac{i}{4\pi}\oint
d\phi\ B\psi^\dag\Gamma^{ij}\psi,
\end{equation}
where $\Gamma^{ij}\equiv\Gamma^{[i}\Gamma^{j]}$ is the
anti-Hermitian $SO(8)$ Lorentz generators acting on spinors, which
we understand as being reduced from $16\times 16$ to $8\times 8$
and acts on positive chirality subspace.

\subsection{Quantization and entropy of near-circular supertubes}

In order to quantize the modes identified before, we now fix the
remaining diffeomorphism. As in the previous section, let us
consider a small deformation from the circular tube with radius
$R_0$ in the 1-2 plane. Then $x^1+ix^2=R(\phi) e^{i\phi}$
with $|R(\phi)-R_0|\,\ll\, R_0$, and $|x^i(\phi)|\,\ll\, R_0$ for all
$i=3,4,\cdots,8$. To identify the quantization conditions, we have
to know the quadratic piece of the gauge-fixed Lagrangian. The
full gauge fixed Lagrangian density becomes
\begin{eqnarray}
{\mathcal{L}}&=&-\sqrt{{\mathcal{L}}^2_b+{\mathcal{L}}^2_f}\nonumber\\
{\mathcal{L}}^2_b&=&B^2-(|\dot{\vec{x}}|^2+\dot{R}^2)
(|\vec{x}'|^2+R^2+(R')^2+B^2)\nonumber\\
&&+(\dot{\vec{x}}\cdot\vec{x}'+\dot{R}R')^2-2EB(\dot{\vec{x}}\cdot\vec{x}'
+\dot{R}R')+(1-E^2)(|\vec{x}'|^2+R^2+(R')^2)\\
{\mathcal{L}}^2_f&=&-2i\left\{B^2-B(\dot{\vec{x}}\cdot\vec{x}'+\dot{R}R')
+(1-E)(|\vec{x}'|^2+R^2+(R')^2)\right\}\psi^\dag\dot{\psi}\nonumber\\
&&-2i\left\{B(|\dot{\vec{x}}|^2+\dot{R}^2)+(E-1)
(B+\dot{\vec{x}}\cdot\vec{x}'+\dot{R}R')\right\}\psi^\dag\psi'\nonumber\,,
\end{eqnarray}
where vectors are in the six dimensional $x^i$ space.
With the choice of the vector potential $A_z=t+a$, the field
strengths are given by $E=1+\dot{a}$, $B=B_0\!-\!a'$. We expand
this action up to quadratic order in $a(\phi,t)$, $r(\phi,t)$,
$x^i(\phi,t)$, $\psi(\phi,t)$. The resulting quadratic Lagrangian density is
\begin{eqnarray}\label{quadratic}
&&{\cal L}_2= \frac{R_0}{B_0}\
\dot{a}\left\{\frac{R_0}{B_0}a'+2r\right\}+\dot{\vec{x}}\cdot\vec{x}'
+\dot{r}r'+iB_0\psi^\dag\dot{\psi}+\frac{B_0^2+R_0^2}{2B_0}
\left\{\dot{r}^2+|\dot{\vec{x}}|^2\right\}
+{R_0^2\over 2 B_0}\dot{a}^2
\,.
\end{eqnarray}
Working within the 1/4 BPS phase moduli space only, the terms of quadratic time
derivative may be dropped since the BPS states are time-independent.

The mode expansion for eight bosonic/fermionic fields are given as
\begin{eqnarray}
\label{boson-mode}
&&A_z=t+a(\phi,t)=t+\sum_{n\neq 0}a_n(t) e^{-in\phi}\ ,\ \
R(\phi,t)=R_0+r(\phi,t)=R_0+\sum_{n\neq 0} r_n(t) e^{-in\phi}\
,\nonumber\\
&&x^i(\phi,t)=x_0^i+\sum_{n\neq 0}x^i_n(t)e^{-in\phi}\ ,\ \
\psi(\phi,t)=\sum_{n\neq 0}\psi_n(t)e^{-in\phi},
\end{eqnarray}
with $a_{-n}=a_n^\dag$, $r_{-n}=r_n^\dag$ and $x^i_{-n}=(x_n^i)^\dag$.
The transverse center of mass positions $x^i_0$ would not affect
the following analysis, so we will neglect them. $\psi(\phi,t)$ and
$\psi_n$'s carry eight components.
Inserting the mode expansions into (\ref{quadratic}) and integrating over
$\phi$ and $z$ (for the zero mode in the $z$ direction), we get
the Lagrangian
\begin{eqnarray}
&& L_2 =2\pi L_z\sum_{n\neq 0}\left[inr_n^\dag\dot{r}_n +
\frac{R_0}{B_0} (in\frac{R_0}{B_0}a_n^\dag + 2r_n^\dag)\dot{a}_n +
in\vec{x}_n^\dag\cdot
\dot{\vec{x}}_n+iB_0\psi_n^\dag \dot{\psi}_n\right]\,.
\end{eqnarray}
Introducing
\begin{equation}
X_{n\pm}\equiv r_n\mp
i\frac{R_0}{B_0}a_n\,,
\end{equation}
the  Lagrangian  becomes
\begin{eqnarray}
&& L_2=2\pi L_z\sum^\infty_{n= 1}\left[i(n+1)X_{n+}^\dag\dot{X}_{n+}
+i(n-1)X_{n-}^\dag\dot{X}_{n-}
 +
2in\vec{x}_n^\dag\cdot
\dot{\vec{x}}_n+2iB_0\psi_n^\dag \dot{\psi}_n\right]\,.
 \label{quant}
\end{eqnarray}
After the Dirac quantization procedure, the resulting commutation
relations read
\begin{eqnarray}
&&[X_{m\pm},X_{n\pm}^\dag]=\frac{1}{2\pi
L_z(n\pm 1)}\,\,\delta_{m,n}\,,\nonumber\\
&&[x^i_m,\,\, x^{j\dag}_n]=\frac{1}{2\pi
L_z(2n)}\,\,\delta_{m,n}\,\,\delta^{i,j}\,,\\
&&\{\psi_m,\,\,\psi_n^\dag\}=\frac{1}{2\pi L_z(2B_0)}\,\,\delta_{m,n}\,,\nonumber
\end{eqnarray}
with all the other commutators vanishing ($i,j=3,4,\cdots, 8$,
$m,n=1,2,\cdots$).
Note that the radius $r_n$ and gauge field $a_n$ modes mix
nontrivially in the commutation relation. Special remark for
$X_{1-}$ is in order: The above relation is meaningless for $X_{1-}$.
This is natural since the dipole deformation of radius $R(\phi)$ is
nothing but the translation of supertube along 1-2 plane~\cite{BHO}.
So we expect that there are true zero modes having the quadratic
time derivative terms only for
their kinetic part.\footnote{In the gravity description, this mode is like
the freedom of translating black holes or supertubes.}
We are not interested in this translational zero mode.

The conserved charges are expressed as
\begin{eqnarray}
&&
J_{12}
=R_0^2+\sum_{n>0}\left\{2 |r_n|^2-
iB_0\psi_{n}^\dag\Gamma_{12}\psi_n\right\}\label{j-expand}, \\
&&
Q_0Q_1=R_0^2+2\sum_{n>0}\!\left\{
\Big|r_n-i\,n\, a_n {R_0 \over B_0} \Big|^2
\!+
\! n^2
|r_n|^2\!+\!
|\vec{x}_n|^2
\!+\!nB_0
|\psi_n|^2\!\right\},
\label{q-expand}
\end{eqnarray}
where $|A|^2=A^\dag A$ is our ordering convention.
The two charges commute, as they should. The first expression
(\ref{j-expand}) determines $R_0^2$ in terms of oscillators and
$J$. Inserting this into (\ref{q-expand}), we obtain
\begin{equation}
\label{charge-osc}
Q_0Q_1\!-\!J=\sum_{n>0}\!\left\{\!n(n\!+\!1)
|X_{n+}|^2\!+\!n(n\!-\!1)
|X_{n-}|^2+
2n^2
|\vec{x}_n|^2
\!+\!2nB_0
|\psi_n|^2\!+\!iB_0\psi_{n}^\dag\Gamma_{12}\psi_n\right\}\,.
\end{equation}
Here we choose a basis for the spinor $\psi$ such that
$i\Gamma_{12}$ is diagonal with four $\pm 1$ eigenvalues with
corresponding modes $\psi_{n\pm}^\alpha$, ($\alpha=1,2,3,4$).
Furthermore, we normalize the oscillators in the canonical way as
follows:
\begin{equation}
X_{n\pm}=\frac{1}{\sqrt{2\pi L_z(n\pm 1)}}Y_{n\pm},\
\vec{x}_n=\frac{1}{\sqrt{2\pi L_z(2n)}}\ \vec{y}_n,\
\psi_{n\pm}^\alpha=\frac{1}{\sqrt{2\pi
L_z(2B_0)}}\Psi_{n\pm}^\alpha
\end{equation}
with $n>0$. The new oscillators satisfy the commutation relations
$[Y_{n\pm},Y_{n\pm}^\dag]=[y_n^i,y_n^{i\dag}]=
\{\Psi_{n\pm}^\alpha,\Psi_{n\pm}^{\alpha\dag}\}=1$. Then we can
rewrite (\ref{charge-osc}) as
\begin{eqnarray}
2\pi L_z(Q_0Q_1-J)&=&\sum_{n>0}\!\left\{\!n
|Y_{n+}|^2\!+n
|Y_{n-}|^2\!+\!n
|\vec{y}_n|^2
\!+\!(n\!+\!\frac{1}{2})
|\Psi_{n+}^\alpha|^2\!+\!
(n\!-\!\frac{1}{2})
|\Psi_{n-}^\alpha|^2\right\}\nonumber\\
&=&\sum_{n>0}\!\left\{\!\sum_{I=1}^8
nN_n^I+\sum_{\alpha=1}^4\left[\!(n\!+\!\frac{1}{2})N_{n+}^\alpha\!+\!
(n\!-\!\frac{1}{2})N_{n-}^\alpha\right]\right\}\ ,\label{number}
\end{eqnarray}
where the last expression contains $8$ classes of bosonic number
operators $N_n^I$ ($I=1,2,\cdots,8$) and $4$ classes of fermionic
number operators $N_{n\pm}^\alpha$ with $\pm$ spins. $2\pi
L_z(Q_0Q_1-J)$ may take half-integer eigenvalues.

The entropy can be counted by considering the generating function
tr\;$(\omega^{2N})=\sum_{n=0}^\infty d_n\omega^n$ where $N$ is the
number operator (\ref{number}), and obtaining the degeneracy $d_n$
with $\frac{n}{2}=2\pi L_z(Q_0Q_1-J)$ being a large half-integer.
One has
\begin{equation}\label{generator}
\mbox{tr}\;(\omega^{2N})=\left(\prod_{m=1}^{\infty}(1-\omega^{2m})\right)^{-8}
\left(\prod_{m=1}^{\infty}(1+\omega^{2m+1})\right)^{4}
\left(\prod_{m=1}^{\infty}(1+\omega^{2m-1})\right)^{4}\ .
\end{equation}
The saddle-point evaluation of the degeneracy
\begin{equation}
\label{contour}
d_n=\frac{1}{2\pi i}\oint
d\omega\frac{\mbox{tr}\;(\omega^{2N})}{\omega^{n+1}},
\end{equation}
requires the behavior of the functions
$f_{\pm}(z)\equiv\prod_{n=1}^{\infty}(1\pm z^n)$ near $z\approx
1^-$. Up to the prefactors that we do not need, we have
\begin{equation}
f_+(z)\sim\exp\left[\frac{\pi^2}{12}\frac{1}{1-z}\right]\ ,\ \
f_-(z)\sim\exp\left[-\frac{\pi^2}{6}\frac{1}{1-z}\right]\ .
\end{equation}
We also note that, as long as we are interested in saddle point
evaluation for large $2\pi L_z(Q_0Q_1-J)$, we may regard
$\omega^{2n\pm 1}$ in (\ref{generator}) as $\omega^{2n}$. Using
the above formulae and noting that $\log \omega \approx
-(1-\omega)$ for $\omega \approx 1^-$, one can see that
(\ref{contour}) gets dominant contributions near
$\log\omega\approx-\sqrt{\frac{\pi^2
(8+4)}{12(n+1)}}\approx-\sqrt{\frac{\pi^2}{n}}$, where $c_B=8$ and $c_F=4$
denotes boson/fermion contributions, respectively. The result is
\begin{equation}
d_n\sim\exp\left[2\pi\sqrt{(c_B+c_F)\frac{n}{12}}\right] =
\exp\left[2\pi\sqrt{n}\right]\ ,\ \ (c_B=2 c_F=8),
\end{equation}
up to the prefactor, which is a suitable power of $n$. Inserting
$n=4\pi L_z(Q_0Q_1-J)$, we get the final expression for the
supertube entropy
\begin{equation}
S=\log(d_n)=4\pi\sqrt{\pi L_z (Q_0Q_1-J)}\ .
\label{DBIe}
\end{equation}
As expected, this is $\sqrt{c_B+c_F}=\sqrt{12}$ times the entropy
(\ref{bos}) from one boson.

One may consider more general case of the multiple
circular supertubes carrying $SO(8)$ Cartans
$J_a=J_{2a\!-\!1,2a}$ for $a=1,2,3,4$. The relevant background is described
by
\begin{equation}
x_{2a-1}+i x_{2a}= R_a e^{i\phi}
\end{equation}
with
\begin{equation}
\Pi_0 B_0=T_2\sum^4_{a=1} R^2_a\,.
\end{equation}
The corresponding angular momentum becomes $J_a=T_2 R_a^2$. Repeating
the above analysis, one may get straightforwardly
\begin{equation}
S=4\pi\sqrt{\pi L_z \left(Q_0Q_1-\sum_{a=1}^4 |J_a|\right)}\,,
\label{DBIee}
\end{equation}
in an appropriate near circular limit.

\section{Microstates in Supermembrane Picture}

In this section we derive the entropy of supertube with charges
$Q_1$, $Q_0$ and angular momentum $J_a (a=1,\cdots,4)$ from the 11-dimensional
M-theory point of view including the contribution from bosons and
fermions. We study equations of motion for a supermembrane with
winding number and momentum along the 11-th direction, which
preserves 1/4 supersymmetry. This approach gives a simple
derivation of BPS equations because fields on the supermembrane
are only 11 bosons $X^M=(t,z,x^i (i=1, \cdots, 8),x^\natural)$ and
a Majorana fermion $\Theta$, which denote the embedding of the
supermembrane into the superspace.

Let us investigate BPS equations for the supermembrane. These are
obtained by analyzing the Killing spinor equations of~\cite{WHN}
\begin{alignat}{3}
&\delta X^M = i \bar{\epsilon} \Gamma^M \Theta + i \bar{\Theta}
\Gamma^M (1+\Gamma) \kappa = 0,\nonumber\\
&\delta \Theta = \epsilon + (1+\Gamma) \kappa = 0,
\end{alignat}
where $\Gamma^M$ are 11-dimensional gamma matrices, and $\Gamma$
is defined as
\begin{alignat}{3}
\Gamma &= \frac{1}{3! \sqrt{-\det P[G(X^M,\Theta)]_{ab}}}
\epsilon^{abc}
\partial_a \Pi^L \partial_b \Pi^M \partial_c \Pi^N \Gamma_{LMN}.
\end{alignat}
Here $P[G(X^M,\Theta)]_{ab}$ is the induced metric on the
world-volume, and $\Pi^M$ are super invariant 1-forms, $\Pi^M =
dX^M - i\bar{\Theta}\Gamma^M d\Theta$. Note that $\Gamma$
satisfies $\Gamma^2=1$ and tr\,$\Gamma=0$. By using the former
property, $\kappa$ can be eliminated and the Killing spinor
equations simply become
\begin{alignat}{3}
&\bar{\Theta} \Gamma^M \epsilon = 0, \label{eq:kilb}
\\
&(1-\Gamma)\epsilon = 0. \label{eq:kilf}
\end{alignat}
Since we are considering the supermembrane which corresponds to
the supertube, the solution should be 1/4
supersymmetric~\cite{HO}:
\begin{alignat}{3}
\epsilon = \frac{1+\Gamma_{t\natural
z}}{2}\frac{1+\Gamma_{t\natural}}{2} \epsilon_0, \label{eq:sol}
\end{alignat}
where $\epsilon_0$ is an arbitrary Majorana spinor.

We then find that Eq.~(\ref{eq:kilb}) has the solution
(\ref{eq:sol}) iff
\begin{alignat}{3}
\Theta &= \frac{1-\Gamma_{t\natural z}}{2}
\frac{1+\Gamma_{t\natural}}{2} \Theta_0, \label{eq:bps1}
\end{alignat}
where $\Theta_0$ is an arbitrary Majorana spinor. We see that
$\Theta$ has 8 real components. It is easy to verify the following
relations:
\begin{alignat}{3}
&\bar{\Theta} \Gamma^z \Theta = \bar{\Theta} \Gamma^i \Theta =
\bar{\Theta} \Gamma_{t\natural} \Theta = \bar{\Theta}
\Gamma_{i\natural} \Theta = \bar{\epsilon} \Gamma_{t\natural}
\Theta = \bar{\epsilon} \Gamma_{z\natural} \Theta = 0,
\nonumber\\
&\Gamma^t \Theta = \Gamma^\natural \Theta = \Gamma^{z\natural}
\Theta = \Gamma_{tz} \Theta.
\end{alignat}

Next we consider Eq.~(\ref{eq:kilf}). The world-volume coordinates
on the supermembrane are identified with $(\tau,\phi,z)$, and we
assume that $x^i$, $x^\natural$ and $\Theta$ depend only on $\tau$
and $\phi$. Then the super invariant 1-forms $\Pi^M$ are expressed
as
\begin{alignat}{3}
\Pi^M &= \dot{\Pi}^M d\tau + {\Pi^M}' d\phi, \quad (M =
t,i,\natural), \notag
\\
\Pi^z &= dz + \dot{\Pi}^z d\tau + {\Pi^z}' d\phi, \label{eq:ansatz}
\end{alignat}
where $\dot{\Pi}^M = \dot{X}^M - i\bar{\Theta}\Gamma^M
\dot{\Theta}$ and ${\Pi^M}' = {X^M}' - i\bar{\Theta}\Gamma^M
\Theta'$. Notations $\dot{X}$ and ${X}'$ represent $\frac{\partial
X}{\partial \tau}$ and $\frac{\partial X}{\partial \phi}$,
respectively. Then $\Gamma$ is written as
\begin{alignat}{3}
&\Gamma = \frac{1}{\sqrt{X}} \Big\{ (\dot{\Pi}^t {\Pi^i}' \!-\!
{\Pi^t}' \dot{\Pi}^i) \Gamma_{tiz} + (\dot{\Pi}^\natural {\Pi^i}'
\!-\! {\Pi^\natural}' \dot{\Pi}^i) \Gamma_{\natural iz} +
(\dot{\Pi}^t {\Pi^\natural}' \!-\! \dot{\Pi}^\natural {\Pi^t}')
\Gamma_{t\natural z} + \dot{\Pi}^i {\Pi^j}' \Gamma_{ijz} \Big\},
\label{eq:gamma}
\\
&\sqrt{X} = \sqrt{- \big( -\dot{\Pi}^{t\,2} + \dot{\Pi}^i{}^2 +
\dot{\Pi}^{\natural\,2} \big) \big( -{\Pi^t}'{}^2 + {\Pi^i}'{}^2 +
{\Pi^\natural}'{}^2 \big) + \big( -\dot{\Pi}^t {\Pi^t}' +
\dot{\Pi}^i {\Pi^i}' + \dot{\Pi}^\natural {\Pi^\natural}'
\big)^2}, \notag
\end{alignat}
where we have defined the volume factor $\sqrt{X}$. From these
equations, we find that Eq.~(\ref{eq:sol}) becomes the solution of
(\ref{eq:kilf}) when
\begin{alignat}{3}
\dot{\Pi}^i = k \, {\Pi^i}', \quad
\dot{\Pi}^t - k{\Pi^t}' = \dot{\Pi}^\natural -
   k{\Pi^\natural}'\,, \label{eq:bps2}
\end{alignat}
where $k$ is a constant. Later we set $k=-1$.
Therefore the BPS equations of the supermembrane corresponding to
the supertube are given by (\ref{eq:bps1}) and (\ref{eq:bps2}).

Now that we have obtained the BPS equations which minimize the
energy of the supermembrane, our next task is to derive conditions
to fix two charges $Q_1$, $Q_0$ and angular momenta.
The two charges are winding number and momentum along the $x^\natural$
direction, so the conditions are written as
\begin{alignat}{3}
\label{q1} Q_1 &= \frac{1}{2\pi} \oint_{-\pi}^\pi d\phi
\frac{{x^\natural}'}{2\pi R_{11}}, \notag
\\
Q_0 &= \oint_{-\pi}^\pi d\phi R_{11} p^\natural, \notag
\\
J_{ij} &= \frac{1}{2\pi} \oint_{-\pi}^\pi d\phi \big( x^ip^j - x^jp^i -
\tfrac{1}{2}S\Gamma^{ij}\Theta \big),
\end{alignat}
where $R_{11}$ is the radius of the 11th circle, and
$p_\natural$ and $S_\alpha$ are the conjugate momenta to
$x^\natural$ and $\Theta^\alpha$, respectively. Note that $2\pi
Q_1$ and $Q_0 L_z$ are integers.
{}From now on we identify $\tau$ and $\phi$ with
$t = 2\pi Q_1 R_{11} \tau$ and
$x^\natural = 2\pi Q_1 R_{11} \phi$, respectively. With this choice, the first
equation in~(\ref{q1}) is trivially satisfied.

To explicitly write down the other conditions, we need the
conjugate momenta. These are calculated from the supermembrane
action:
\begin{alignat}{3}
S_{\text{M2}} &= S_{\text{NG}} + S_{\text{WZ}}, \notag
\\
S_{\text{NG}} &= -T_2 \int d\tau d\phi dz \sqrt{X}, \label{eq:M2act}
\\
S_{\text{WZ}} &= T_2 \int d\tau d\phi dz \big( i{x^\natural}'
\bar{\Theta} \Gamma^t \dot{\Theta} -i \dot{x}^\natural \bar{\Theta} \Gamma^t
\Theta' + i \dot{t} \bar{\Theta} \Gamma^t \Theta' \big)\,.
\notag
\end{alignat}
We have reduced the degrees of freedom of
$\Theta$ by using (\ref{eq:bps1}), and assumed the same ansatz as
(\ref{eq:ansatz}). By imposing the BPS condition~(\ref{eq:bps2}),
the conjugate momenta are given by
\begin{alignat}{3}
p^i &= - T_2 {x^i}' ,
\label{eq:constr1}
\\
p^\natural &= \frac{T_2 |\vec{x}'|^2}{{x^\natural}'} - 2iT_2
\bar{\Theta}\Gamma^t \Theta',
\label{eq:mom}
\\
S &= 2i T_2 {x^\natural}' \bar{\Theta} \Gamma^t.
\label{eq:constr2}
\end{alignat}
The conditions for fixing charges and angular momenta are then
written as
\begin{alignat}{3}
\frac{Q_1Q_0}{T_2} &= \frac{1}{2\pi} \oint d\phi \big(
|\vec{x}'|^2 - 2i {x^\natural}' \bar{\Theta} \Gamma^t \Theta'
\big), \notag
\\
\frac{J_a}{T_2} &= \frac{1}{2\pi} \oint d\phi \big(- x^{2a\!-\!1} {x^{2a}}' +
x^{2a}{x^{2a\!-\!1}}' -i{x^\natural}' \bar{\Theta} \Gamma^t
\Gamma^{2a\!-\!1,2a}\Theta\big)\,, \label{eq:MJ}
\end{alignat}
where $a=1,2,3,4$ labels the $SO(8)$ Cartans.
These are the constraints obtained by fixing $Q_1$, $Q_0$ and
angular momenta.

Now let us consider the quantization of $x^i$ and $\Theta$.
{}From the BPS equations $(\partial_\tau + \partial_\phi) x^i = 0$,
$x^i$ contain only right moving modes,
\begin{alignat}{3}
x^i(\tau,\phi) = x^i_0 + \frac{p^i_0}{2\pi L_zT_2}(\tau-\phi) +
\frac{i}{\sqrt{4\pi T_2 L_z}} \sum_{m\neq 0} \frac{\sqrt{|m|}}{m}
x^i_m e^{-im(\tau-\phi)}.
\end{alignat}
Since $x^i$ and $p^i$ are related as
(\ref{eq:constr1}), we need the Dirac quantization of the
constrained system. After some calculations, we obtain the
commutation relations $[x^i(\phi), p^j(\phi')] = \frac{i}{2L_z}
\delta^{ij} \delta(\phi-\phi')$, and hence
\begin{alignat}{3}
[x^i_m,x^{j\dagger}_n] &= \delta^{ij} \delta_{mn}.
\end{alignat}
The Majorana fermion $\Theta$ is also treated similarly.
{}From the equations of motion obtained by (\ref{eq:M2act}) and
BPS equations (\ref{eq:bps2}), we only need the right  moving modes
and $\Theta_\alpha (\alpha = 1,\cdots,8)$ are expanded as
\begin{alignat}{3}
\Theta_\alpha &= \frac{1}{\sqrt{8\pi T_2 {x^\natural}' L_z}}
\sum_m \Theta_{m \alpha} e^{-im(\tau-\phi)}.
\end{alignat}
Since $\Theta_\alpha$ and $S^\alpha$ are related as
(\ref{eq:constr2}), after the calculation of Dirac brackets we
obtain $\{ \Theta_\alpha, S^\beta \} = \frac{i}{2L_z}
\delta_\alpha^\beta \delta(\phi-\phi')$, and hence
\begin{alignat}{3}
\{ \Theta_{m \alpha}, \Theta_n^{\dagger \beta} \} =
\delta_\alpha^\beta \delta_{mn},
\end{alignat}
where $\Theta^\dagger = -i \Theta^T C^{-1} \Gamma^t$.

After the quantization, the constraints (\ref{eq:MJ}) reduce to
\begin{alignat}{3}
2\pi L_z Q_0 Q_1 &= \sum_{i=1}^8 \sum_{m=1}^\infty m
x^{i\,\dagger}_m x^i_m + \sum_{\alpha=1}^8 \sum_{m=1}^\infty m
\Theta_m^{\dagger \alpha} \Theta_{m \alpha}, \notag
\\
2\pi L_zJ_a &= i \sum_{m=1}^\infty \big( x^{2a\!-\!1}_m x^{2a\dagger}_m -
x^{2a}_m x^{2a\!-\!1\dagger}_m \big) - i \sum_{\alpha=1}^8 \sum_{m=1}^\infty
\frac{1}{2} \Theta_m^{\dagger \alpha} \Gamma^{2a\!-\!1,2a} \Theta_{m
\alpha}, \label{eq:const}
\end{alignat}
Note that $2\pi L_z Q_0 Q_1$ is an integer. The volume of the
phase moduli space is obtained exactly by counting the
configurations of $x^i$ and $\Theta_\alpha$ which satisfy the
above constraints.

We choose the spinor eigen basis digonalizing
$i\Gamma^{2b\!-\!1,2b}$ ($b=1,2,3$). One has
\begin{equation}
i\Gamma^{2b\!-\!1,2b} \Theta_{n,\vec{s}}=s_b \Theta_{n,\vec{s}}\,,
\end{equation}
where $\vec{s}=(s_1,s_2,s_3)$ with $s_b=\pm 1$. Since $\prod^4_{a=1}
\Gamma^{2a\!-\!1,2a}\Theta_{n,\vec{s}}=\Theta_{n,\vec{s}}$,
the eigenvalue $s_4$ of $i\Gamma^{78}$ is given by $s_1 s_2 s_3$.

The combination of the constraints in (\ref{eq:const}) give
\begin{alignat}{3}
N &= 2\pi L_z (Q_0 Q_1 - \sum_a J_a) \notag
\\
&= \sum_{m=1}^\infty \bigg\{  \sum^4_{a=1}
\left( (m+1)A_{am}^\dagger A_{am} + (m-1)
B_{am}^\dagger B_{am}\right) +
\sum_{\vec{s}}\Big(m+{1\over 2}{\sum^4_{a=1}s_a} \Big)
\Theta_{m\vec{s}}^{\dagger} \Theta_{m \vec{s}} \bigg\},
\label{const1}
\end{alignat}
where we have defined
\begin{alignat}{3}
&A_{am} = \frac{1}{\sqrt{2}}(x^{2a\!-\!1}_m + i x^{2a}_m), \qquad & &B_{am} =
\frac{1}{\sqrt{2}}(x^{2a\!-\!1}_m - i x^{2a}_m),
\\
&[A_{am},A_{bm}^\dagger] = \delta_{ab},
\qquad & &[B_{am},B_{bm}^\dagger] = \delta_{ab}. \notag
\end{alignat}
The second constraint in (\ref{eq:const}) is written as
\begin{eqnarray}
\sum_{m=1}^\infty
(B_{am}^\dag B_{am} - A_{am}^\dag A_{am}) -
\frac{1}{2} \sum_{\vec{s}} \sum_{m=1}^\infty s_a \Theta_{m\vec{s}}^{\dagger}
\Theta_{m\vec{s}}
= 2\pi L_z J_a.
\label{const2}
\end{eqnarray}
We can consider this determines $B_{a1}$ mode which is absent from
(\ref{const1}). Thus the number of microstates can be counted by
taking only the constraint~(\ref{const1}) into account.

We consider the case $N\gg 1$. As in the previous sections,
let us compute the partition function
\begin{alignat}{3}
G(w) = \text{tr}\, w^N &= \sum_{n=0}^\infty d_n w^n \notag \\
&= \frac{2(1+w^{-1})(1-w)^4}{(1+w)(1+w^2)}
\prod_{m=1}^\infty \frac{(1+w^m)^8}{(1-w^m)^8} \notag \\
&= \frac{(1-w)^4(1+w^{-1})}{2^3(1+w)(1+w^2)}
\frac{\vartheta_{10}^4(0,\tau)}{\eta^{12}(\tau)}  \notag
\\
&= \frac{(1-w)^4(1+w^{-1})}{2^3(1+w)(1+w^2)} \Big(- \frac{\ln w}{2\pi} \Big)^4
\frac{\vartheta_{01}^4(0,-1/\tau)}{\eta^{12}(-1/\tau)} ,
\label{mod2}
\end{alignat}
where we have used the eta function defined in Eq.~(\ref{eta}) and
\begin{alignat}{3}
&\vartheta_{10}(0,\tau) = 2 w^{\frac{1}{8}} \prod_{m=1}^\infty
(1-w^m)(1+w^m)^2, \notag
\\
&\vartheta_{01}(0,\tau) = \prod_{m=1}^\infty
(1-w^m)(1-w^{m-1/2})^2,
\end{alignat}
with $w = e^{2\pi i\tau}$, and the modular transformations of the
eta (\ref{modular}) and
\begin{alignat}{3}
\vartheta_{10}(0,-1/\tau) = \sqrt{-i\tau} \,
\vartheta_{01}(0,\tau).
\end{alignat}
The result (\ref{mod2}) gives the asymptotic formula
\begin{equation}
G(w) \sim \pi^4 \Big( -\frac{\log w}{2\pi} \Big)^8
\exp\left(-\frac{2\pi^2}{\log w}\right), \label{asym1}
\end{equation}
for $w \sim 1^-$. The saddle point approximation enables us to derive
the final result for the degeneracy for $N=n$ as
\begin{alignat}{3}
d_n = \frac{1}{2\pi i} \oint \frac{G(w)}{w^{n+1}} dw \sim e^{2\pi\sqrt{2n}}.
\end{alignat}
In this way we obtain the entropy
\begin{alignat}{3}
S = \log d_N 
\sim 4\pi
\sqrt{\pi L_z \left(Q_0Q_1-\sum_{a=1}^4|J_a|\right)},
\label{mbrane}
\end{alignat}
in agreement with (\ref{DBIee}).

We would like to emphasize that we have not used the near circular
condition $(Q_0 Q_1-J)/J \, \ll \, 1$ anywhere for the evaluation
of the entropy in this section. So the entropy formula (\ref{mbrane})
will be valid beyond the near circular limit.
Let us confirm this for the case of $\Theta=0$ and $J_a=0~(a=2,3,4)$
for simplicity. The expectation value of the radius squared is given by
\begin{alignat}{3}
R^2 \equiv \frac{1}{2\pi} \oint d\phi |\vec{x}|^2
= \frac{1}{2\pi L_z T_2} \sum_{m=1}^\infty \sum_{a=1}^4
\Big( \frac{1}{m} A_{a m}^\dagger A_{a m}
+ \frac{1}{m} B_{a m}^\dagger B_{a m} \Big).
\label{r}
\end{alignat}
As discussed in the introduction, we need to have $R_S \ll R$ for
the validity of our counting of microstates. The value~(\ref{r}) of $R$
should be estimated under the constraints~(\ref{eq:const})
or (\ref{const1}) and (\ref{const2}). The constraint~(\ref{const2})
tells us that we must excite certain amount of $B_{am}$ and $A_{am}$,
and $R$ is smaller if those with larger $m$ are excited, but there
is an upper limit on possible $m$ from the first equation in
(\ref{eq:const}). We thus find that the minimum of $R$ is attained
when we excite $2\pi L_z J_1$ of $B_{1m}$ for $m\sim Q_0Q_1/J_1$, giving
\begin{alignat}{3}
R \sim  \sqrt{\frac{J_1}{Q_0Q_1}} \sqrt{\frac{J_1}{T_2}}.
\end{alignat}
Thus, as long as $J_1/Q_0Q_1$ is not very small,
$R_S \ll R$ can be satisfied for large $J_1$ and therefore
the entropy formula (\ref{mbrane}) is valid.

\section{Conclusions}

In this paper we have presented two approaches to the counting
of the number of microstates for supertubes specified by the F1
and D0 charges and angular momenta, and derived consistent entropy
formula.

There are corresponding supergravity microstates and, thus, we
count the degeneracy of the geometries with the asymptotic
geometry and charges fixed. The correspondence demonstrates the
existence of the quantized microstates specified by the
distinguishable supergravity fields. Thus although we do not know
how to do precisely, there must be a clear way to sum over
geometries with an appropriate measure. This has been suspected in
many cases including the thermal AdS/CFT
correspondence~\cite{witten}, where one has competing
contributions from the AdS Schwarzschild black hole and the
Euclidean AdS geometry of temporal circle size related to the
inverse temperature.

Indeed the related black hole entropy may be understood from the
microstates. Since the horizon area of the supergravity supertubes
are zero in any cases, the situation here is rather confusing.
However, the proposal of Sen~\cite{sen} may be applied and the
stretched horizon area of the rotationally symmetric solution may
be shown to agree to the entropy~\cite{mathur}.

The situation of D1-D5-P~\cite{lunin} which is related to F1-D0-D4
by a U-duality is different. [D1 (5)-D5 (56789)-P (5) where the
numbers in the parenthesis represent momentum direction or
extending directions, is related to F1 (5)-D0-D4 (6789) by the
successive transformations of S, T5, S, T56789.] The rotationally
symmetric black hole solution of D1-D5-P has a nonvanishing
horizon area and the corresponding entropy can be explained by the
CFT counting~\cite{strovafa}.

When we add $J_{12}$ angular momentum to the F1 (5)-D0-D4
(6789)-D2 (5$\theta_{12}$), the configuration describes the
supertubes intersecting with D4-branes, which preserve four real
supersymmetries. By the same U-duality transformation, the above
is related to the D1 (5)-D5 (56789)-P (5)-KK5
(6789$\theta_{12}$)~\cite{luninmal} where $\theta_{12}$ represents
that the KK monopole or the D2 form a curve in the (12) plane.
Similarly $J_{34}$ may be added too.

Since the supertube ending on D4 in
Refs.~\cite{peet,baklee,kimlee,lambert} has angular momenta in (6789)
plane only, e.g. F1 (5)-D0-D4 (6789)-D2 (5$\theta_{67}$), the
above configurations of the curve in (1234) plane are different in
their expansion directions of D2 and have not been found in the
field theory description.

Considering the supertubes suspended between two D4-branes of
large separations, the corresponding entropy is expected as
$S={4\pi \over \sqrt{2}} \sqrt{\pi L(Q_0 Q_1 -\sum |J|)}$, where
the sum is over the $SO(4)$ Cartans in (6789) plane. The curve
cannot escape to the (1234) plane because the supertube ends on D4-branes.
Thus only four arbitrary bosonic fluctuations remain.
Furthermore they preserve four real supersymmetries and the number
of arbitrary fermionic fluctuations should be reduced to four.
Hence one has the $1/\sqrt{2}$ factors. (The half factor for the
numbers of degrees goes inside of the square root.)

For many D4-branes, the above formula would have a straightforward
generalization. For the supertubes connecting D4-branes, one may
have in principle five independent charges; D0, F1, D4 and two
Cartans of the angular momenta.

For these cases, one has clean examples of the gravity
microstates, black hole solutions whose horizon area reproduces
the entropy, and the corresponding filed theory description of the
microstates. But the detailed and complete construction awaits
more endeavors.

Finally, the formula for the cross sectional area ${\cal
A}\,\sim\, g_s \ell_s^2 N_0 N_1$ in (\ref{areazero}) is reminiscent
of the quantum foam in Ref.~\cite{vafa}. Since the counting and
the partition function may be also related to the black hole
partition function~\cite{stro}, there seems to be some connections
of the microstates to the quantum foam in Ref.~\cite{vafa}. Any
clue in this direction will be very interesting.

\vspace{.7cm} \noindent{\large\bf Acknowledgments}

YH would like to thank H. Kajiura, N. Sasakura and S. Sugimoto.
DB is supported in part by KOSEF ABRL R14-2003-012-01002-0 and
KOSEF R01-2003-000-10319-0. The work of YH was supported in part
by a Grant-in-Aid for JSPS fellows. SK is supported in part by
BK21 project of the Ministry of Education, Korea, and a 2003
Interdisciplinary Research Grant of Seoul National University. NO
was supported in part by Grants-in-Aid for Scientific Research
Nos. 12640270 and 02041.

\end{document}